# Exponential Steepest Ascent from Valued Constraint Graphs of Pathwidth Four

**Artem Kaznatcheev**
Department of Mathematics, and Department of Information and Computing Sciences, Utrecht University, Utrecht, The Netherlands

**Melle van Marle**
Department of Mathematics, and Department of Information and Computing Sciences, Utrecht University, Utrecht, The Netherlands

## Abstract

We examine the complexity of maximising fitness via local search on valued constraint satisfaction problems (VCSPs). We consider two kinds of local ascents: (1) steepest ascents, where each step changes the domain that produces a maximal increase in fitness; and (2) $\prec$-ordered ascents, where – of the domains with available fitness increasing changes – each step changes the $\prec$-minimal domain. We provide a general padding argument to simulate any ordered ascent by a steepest ascent. We construct a VCSP that is a path of binary constraints between alternating 2-state and 3-state domains with exponentially long ordered ascents. We apply our padding argument to this VCSP to obtain a Boolean VCSP that has a constraint (hyper)graph of arity 5 and pathwidth 4 with exponential steepest ascents. This is an improvement on the previous best known construction for long steepest ascents, which had arity 8 and pathwidth 7.

## 1 Introduction

Local search is often used in combinatorial optimisation. One of the most common methods for choosing which local modification to make is the steepest ascent algorithm, which at each step selects the highest-value option from the neighbours of the current state. Clearly such an algorithm could get trapped at local optima that might prevent it from reaching a higher optimum. Surprisingly, Johnson, Papadimitriou and Yannakakis [8] showed that for problems that are hard for the complexity class of polynomial local search (PLS), even local optima can be intractable to find – regardless of what polynomial time algorithm is used for the search. As such, it is natural to ask: under what conditions could popular local search algorithms like steepest ascent be guaranteed to find even a local optimum in reasonable time? Or stated in term of intractability: for what problems does steepest ascent not find a local optimum quickly, taking instead an exponential number of steps before arriving at any local optimum.

Many combinatorial optimisation problems can be formulated as valued constraint satisfaction problems (VCSPs). Since weighted 2-SAT is PLS-complete [13] and a special case of binary Boolean VCSPs, it is believed to be intractable to find local optima in general VCSPs. It is also possible to create VCSPs where every ascent from some initial assignment is exponentially long. VCSPs of bounded treewidth, however, are tractable – even for finding global optima – by using a non-local-search algorithm [1, 2]. But the existence of efficient non-local algorithm does not mean that local search algorithms will find optima efficiently.

Even in the case of VCSPs of bounded treewidth, a (reasonable) local search algorithm like steepest ascent may take a long time to equilibrate. Cohen et al. [3] have provided a Boolean VCSP with arity-8 and pathwidth 7 (and treewidth 7), on which an exponentially long steepest ascent exists. This shows that what is tractable/intractable for steepest ascent is distinct from what is tractable/intractable for non-local algorithms. We currently do not yet have a full characterisation of the intractability class for steepest ascent. In this article, take a next step towards this full characterisation and lower the threshold for intractability, by constructing a VCSP of arity-5 and pathwidth-4 that has exponentially long steepest ascents. We do this in four steps:

**Section 3:** Introduce a general padding argument that allows us to simulate any ascent respecting an ordering of the domains by a steepest ascent (see Theorem 8).
**Section 4:** Introduce a new VCSP (specifically path made by alternating two different 2-by-3 constraints between variables of domain size alternating between 2 and 3) that produces an exponentially long ordered ascent (see Proposition 11).
**Section 5:** Apply the padding argument to the 2-by-3 construction. This yields a 3-by-5 construction, implementable by a VCSP with ternary constraints, on which the original ordered ascent is simulated by a steepest ascent.
**Section 6:** Encode the expanded domains using Boolean variables, and apply some tricks to get the resulting VCSP to arity 5 and pathwidth 4 while preserving the exponentially long steepest ascent.

## 2 Background

Let $D$ be a set and $R \subset D \times D$ a binary relation on $D$. We call $(D, R)$ a *domain* and $R$ the *transition relation*. We omit $R$ when it is a complete graph or obvious from context.

We will consider local search problems on *search spaces* of the form $D_1 \times D_2 \times \cdots \times D_n$, where the $(D_i, R_i)$ are domains. We call the elements of $D_1 \times \cdots \times D_n$ *assignments*. We consider two assignments $x, y \in D_1 \times \cdots \times D_n$ to be as adjacent iff $x$ differs from $y$ at exactly one position, say $k$, and $(x_k, y_k) \in R_k$. We view $R_k$ as being undirected, so that $(x_k, y_k) \in R_k$ implies that both the transition from $x_k$ to $y_k$ and the transition from $y_k$ to $x_k$ are allowed. $N(x)$ is the set of assignments adjacent to $x$.

▶ **Definition 1.** *Let $D_1 \times \cdots \times D_n$ be a search space, and $f : D_1 \times \cdots D_n \to \mathbb{Z}$ a function. We call $f$ a* fitness function *and the pair $(D_1 \times \cdots \times D_n, f)$ a* fitness landscape.[1]

We can represent fitness landscapes using a collection of constraints. A *valued constraint* on $D_1 \times \cdots \times D_n$ with scope $S \subseteq [n]$ is a function $C_S : \prod_{i \in S} D_i \to \mathbb{Z}$. the size $|S|$ of the scope is the *arity* of the constraint.

In general, we can represent a constraint of arity $n$ by an $n$-dimensional tensor, whose fibers are indexed by the domains in the scope of the constraint. In particular, this means that a binary constraint between two domains $D_k$ and $D_l$ can be represented by a matrix whose rows are indexed by $D_k$ and whose columns are indexed by $D_l$ (or vice versa).

▶ **Example 2.** Let $C_{\{i,j\}}$ be a binary constraint between domains $D_i = \{u_1, u_2, u_3, u_4\}$ and

[1] We could have used the more traditional 'value' or 'reward', but we prefer 'fitness' given the connection to biological evolution that we discuss in Section 7.

$D_j = \{v_1, v_2, v_3\}$, then we can represent $C_{\{i,j\}}$ by

$$C_{\{i,j\}} = \begin{pmatrix} C_{\{i,j\}}(u_1,v_1) & C_{\{i,j\}}(u_1,v_2) & C_{\{i,j\}}(u_1,v_3) \\ C_{\{i,j\}}(u_2,v_1) & C_{\{i,j\}}(u_2,v_2) & C_{\{i,j\}}(u_2,v_3) \\ C_{\{i,j\}}(u_3,v_1) & C_{\{i,j\}}(u_3,v_2) & C_{\{i,j\}}(u_3,v_3) \\ C_{\{i,j\}}(u_4,v_1) & C_{\{i,j\}}(u_4,v_2) & C_{\{i,j\}}(u_4,v_3) \end{pmatrix} \tag{1}$$

▶ **Definition 3.** *(Following [7]) Let $\mathbb{F} = (D_1 \times \cdots \times D_n, f)$ be a search space, and let $\mathcal{C} = \{C_{S_1}, C_{S_2}, \ldots, C_{S_m}\}$ be a set of valued constraints on $D_1 \times \cdots \times D_n$. We say that $\mathcal{C}$ implements $\mathbb{F}$ if for all $x \in D_1 \times \cdots \times D_n$, we have $f(x) = \sum_{i=1}^{m} C_{S_i}((x_j)_{j \in S_i})$. An assignment $x$ is called a* local solution *if $f(x) \geq f(y)$ for all $y \in N(x)$. The goal is to find a local solution.*

To any VCSP $\mathcal{C}$ on $n$ variables, we can associate a *constraint (hyper)graph*, whose set of vertices is given by $[n]$. For each constraint $C_s$, there is a (hyper)edge $S$, labeled by $C_s$.

Note that when the value of a single variable changes within a VCSP, the accompanying change in fitness value is determined entirely by those constraints whose scope contains this variable. Because of this, it is often useful to consider the "restricted" fitness function resulting from only considering these domains. Let $f$ be a fitness function associated to some VCSP $\mathcal{C}$, and let $\mathcal{C}[k] \subset \mathcal{C}$ be the set of constraints whose scope contains the $k$-th variable. We use $f_k$ to denote the function given by $f_k(u; z) = \sum_{S_i \in \mathcal{C}[k]} C_{S_i}(u, (z_j)_{j \in S_i \setminus \{k\}})$ where $u$ is the value of the $k$-th variable, and $z$ is a (sub-)assignment consisting of values for all variables who share a constraint with the $k$-th variable.

▶ **Definition 4.** *Let $\mathbb{F} = (D_1 \times \cdots \times D_n, f)$ be a fitness lanscape, and let $p = (x^t)_{t=0}^{T} \subset (D_1 \times \cdots \times D_n)^T$ be a sequence of assignments in $D_1 \times \cdots \times D_n$. We call $p$ an* ascent *on $\mathbb{F}$ if for all $t < T$, we have $x^{t+1} \in N(x^t)$, we have $f(x^t) < f(x^{t+1})$, and $x^T$ is a local solution.*

In this article, we are focused on ascents which take steps that most increase fitness.

▶ **Definition 5.** *Let $\mathbb{F} = (D_1 \times \cdots \times D_n, f)$ be a fitness landscape, and let $p = (x^t)_{t=0}^{T}$ be an ascent on $\mathbb{F}$. We call $p$ a* steepest ascent *on $\mathbb{F}$ if for all $t < T$ and for all $y \in N(x^t)$, we have $f(y) \leq f(x^{t+1})$.*

In other words, at any step, all neighbours of an assignment in $p$ have fitness less than or equal to the fitness of the next assignment in $p$.

We will use the following notation to represent local changes: if $y$ can be generated from $x$ by changing the $k$-the entry from $x_k = u$ to $y_k = v$, we write $y = x[k : v]$.[2] Furthermore, if $p = (x^t)_{t=0}^{T}$ is an ascent, we write $u \xrightarrow{k}_{p(t)} v$ to denote that the transition from $x^t$ to $x^{t+1}$ in $p$ consists of replacing symbol $u$ with symbol $v$ at position $k$. If the ascent is clear from context, we drop the $p$ from the notation.

▶ **Definition 6.** *Let $\mathbb{F} = (D_1 \times \cdots \times D_n, f)$ be a fitness landscape, and let $p = (x^t)_{t=0}^{T}$ be an ascent on $\mathbb{F}$. Let $\prec$ be an ordering of the indices. We call $p$ a* $\prec$-ordered ascent *on $\mathcal{C}$ if the following holds. For any $t < T$, if $x^{t+1} = x^t[k : v]$, then, for all $j \prec k$ and for all $u \in D_j$ with $(x_j^t, u) \in R$, we have $f(x^t[j : u]) \leq f(x^t)$.*

In other words, at any step, the ascent changes an entry in the domain with $\prec$-minimal index where a change can yield a fitness increase.

[2] Note that $y = x[k : v]$ is equivalent to $x = y[k : u]$.

# 3 Steepest Ascent Simulation of Ordered Ascents

Given an ordered ascent on some fitness landscape, we show how to construct a new fitness landscape that "simulates" the ordered ascent with a steepest ascent. This will be done by expanding the domain and then encoding the expanded domain using Boolean variables.

## 3.1 Domain Expansion

Let $\mathbb{F} = (D_1 \times \cdots \times D_n, f)$ be a fitness landscape. Let $p = (x^t)_{t=0}^T$ be a $\prec$-ordered ascent on $\mathbb{F}$. For any $k \in \{1, 2, \ldots, n\}$, we expand $D_k$ by adding *intermediate states* $\sigma_{uv} = \sigma_{vu}$ for all $(u, v) \in R_k$. We call the elements of the original domain $D_k$ *main states*. We denote the resulting expanded domain by $\widehat{D}_k \supset D_k$. We define the new transition relation on the expanded domains to be $\widehat{R}_k = \{(u, \sigma_{uv}) \mid (u, v) \in R_k\} \cup \{(\sigma_{uv}, v) \mid (u, v) \in R_k\}$. This relation ensures that the only possible transition are those from a main state to an intermediate state and vice versa.

We now construct a new fitness function $\widehat{f} : \widehat{D}_1 \times \cdots \times \widehat{D}_n \to \mathbb{Z}$. For any $x \in D_1 \times \cdots \times D_n$ – i.e., any $x$ containing only main states – we set:

$$\widehat{f}(x) := (2n + 1) f(x) \qquad \text{if } x \text{ contains only main states.} \tag{2}$$

For an assignment $x$ containing a single intermediate state $\sigma_{uv}$ at position $k$ and main states at all other positions, if $f(x[k : u]) \neq f(x[k : v])$ then we set $\widehat{f}(x)$ to be:

$$\widehat{f}(x) := n - k + 1 + (2n + 1) \min_{w \in \{u,v\}} f(x[k : w]) \qquad \text{if } x \text{ containts exactly one intermediate state } \sigma_{uv} \text{ at position } k. \tag{3}$$

If $f(x[k : u]) = f(x[k : v])$ then we set $\widehat{f}(x) = (2n + 1) \min_{w \in \{u,v\}} \{f(x[k : w])\}$.

Next, let $x$ be an assignment that contains exactly two intermediate symbols $\sigma_{u_j v_j}, \sigma_{u_k v_k}$ at positions $j$ and $k$ respectively. We do not want such an assignments to appear in the steepest ascent. To ensure this, we require that $\widehat{f}$ satisfies the following:

$$\widehat{f}(x) \leq 2n - (j + k) + 2 + (2n + 1) \min_{\substack{w \in \{u_j, v_j\} \\ w' \in \{u_k, v_k\}}} f(x[j, k : w, w']) \qquad \text{if } x \text{ contains exactly two intermediate states } \sigma_{u_j v_j} \text{ and } \sigma_{u_k v_k} \text{ at positions } j \text{ and } k. \tag{4}$$

For remaining assignments with more than two intermediate symbols, $\widehat{f}$ may take any value.

## 3.2 Steepest Ascent Simulation

▶ **Definition 7.** *Given an ascent $p = (x^t)_{t=0}^T$ on a fitness landscape $(D_1 \times \cdots \times D_n, f)$, we define an ascent $\widehat{p} = (\widehat{x}^t)_{t=0}^{2T}$ on the new fitness landscape $(\widehat{D}_1 \times \cdots \times \widehat{D}_n, \widehat{f})$ as:*

$$\widehat{x}^t = \begin{cases} x^s, & t = 2s; \\ x^s[k : \sigma_{uv}], & t = 2s + 1, \ x^{s+1} = x^s[k : v] \text{ and } x^s = x^{s+1}[k = u]. \end{cases} \tag{5}$$

*Since $\widehat{p}$ alternates between main states of $p$ and the relevant intermediate states between them, we say that $\widehat{p}$* simulates *$p$.*

▶ **Theorem 8.** *Let $\mathbb{F} = (D_1 \times \cdots \times D_n, f)$ be a fitness landscapes, and let $\prec$ be an ordering on $\{1, 2, \ldots, n\}$. Suppose $p$ is a $\prec$-ordered ascent on $\mathbb{F}$ and $\widehat{p}$ simulates $p$. Then, $\widehat{p}$ is a steepest ascent on $(\widehat{D}_1 \times \cdots \times \widehat{D}_n, \widehat{f})$.*

By reindexing the domains, we may assume that $p$ is $<$-ordered without loss of generality. We prove this theorem through the following two lemmas. The first shows that the transitions from main states into intermediate states are steepest ascent steps. The second shows that the transitions from intermediate states into main states are steepest ascent steps.

▶ **Lemma 9.** *Let $x^t, x^{t+1} \in p$, with the transition between these two states being $u \xrightarrow{k}_{p(t)} v$. Then, the highest fitness neighbour of $\widehat{x}^{2t} \in \widehat{p}$ is $\widehat{x}^{2t}[k : \sigma_{uv}]$. Moreover, $\widehat{x}^{2t}[k : \sigma_{uv}]$ has higher fitness than $\widehat{x}^{2t}$.*

**Proof.** We begin by noting that $\widehat{x}^{2t}$ contains only main states. Due to the nature of the encoding, the only possible transitions are those flipping a main state to an intermediate state. Note that by equation (3), we have

$$\widehat{f}(\widehat{x}^{2t}[k : \sigma_{uv}]) - \widehat{f}(\widehat{x}^{2t}) = n - k + 1 > 0. \tag{6}$$

Consider any $l < k$. Since we may assume $p$ is a $<$-ordered ascent, we know that for any neighbour $x^t[l : w]$ of $x^t$, we have $f(x^t[l : w]) \leq f(x^t)$. It follows that

$$\widehat{f}(\widehat{x}^{2t}[l : \sigma]) \leq \widehat{f}(\widehat{x}^{2t}) < \widehat{f}(\widehat{x}^{2t}[k : \sigma_{uv}]$$

for any intermediate state $\sigma$ in $\widehat{D}_l$.

Next, consider any $l \geq k$. From equation (3), it is clear that the fitness increase from a flip to an intermediate state will be at most $n - l + 1 \leq n - k + 1$. Thus, the transition from $\widehat{x}^{2t}$ to neighbour $\widehat{x}^{2t}[k : \sigma_{uv}]$ with fitness increase of $n - k + 1$, is a steepest step as desired. ◀

▶ **Lemma 10.** *Let $x^t, x^{t+1} \in p$, with the change between these two states being $u \xrightarrow{k}_{p(t)} v$. Then, the highest fitness neighbour of $\widehat{x}^{2t+1} \in \widehat{p}$ is $\widehat{x}^{2t+1}[k : v]$. Moreover, $\widehat{x}^{2t+1}[k : v]$ has higher fitness than $\widehat{x}^{2t+1}$.*

**Proof.** We begin by noting that by definition, $\widehat{x}^{2t+1}$ consists of the intermediate state $\sigma_{uv}$ at position $k$, and main states at all other positions. We have $\widehat{f}(\widehat{x}^{2(t+1)}) - \widehat{f}(\widehat{x}^{2t}) \geq 2n + 1$, since $\widehat{x}^{2(t+1)} = x^{t+1}$, $\widehat{x}^{2t} = x^t$ and $f(x^{t+1}) > f(x^t)$. Moreover $\widehat{f}(\widehat{x}^{2t+1}) - \widehat{f}(\widehat{x}^{2t}) = n - k + 1$. Thus $\widehat{f}(\widehat{x}^{2(t+1)}) - \widehat{f}(\widehat{x}^{2t+1}) \geq 2n + 1 - (n - k + 1) \geq n + 1$. Since any transition to an intermediate state can yield a fitness increase of at most $n$, we know that such a transition will never be preferred over the transition from intermediate state $\sigma_{uv}$ to main state $v$, which yields a fitness increase of at least $n + 1$. By definition of the transition relation, the only alternative transition is from $\sigma_{uv}$ to main state $u$ but this decreases fitness by $n - k + 1$. Thus the transition from $\widehat{x}^{2t+1}$ to neighbour $\widehat{x}^{2t+1}[k : v]$ is steepest step as desired. ◀

Together Lemmas 9 and 10 imply that $\widehat{p}$ is a steepest ascent.

## 3.3 Boolean Encoding of Expanded Domains

Note that we can encode each $(\widehat{D}_k, \widehat{R}_k)$ using $|D_k|$ Boolean variables. Without loss of generality, we may assume that $D_k = \{1, 2, \ldots, |D_k|\}$. We can now encode any main state $u \in D_k$ by the string of length $|D_k|$ that contains a 1 at position $u$ and a 0 in all other positions. Any intermediate state $\sigma_{uv}$ can likewise be encoded by the string of length $|D_k|$ containing a 1 at positions $i$ and $j$ and a 0 at all other positions.

Using this encoding we ensure that all main states are more than a single bit-flip away from each other. At the same time, the intermediate states are exactly one flip away from the corresponding two main states (and more than one flip away from all other main states). Thus, the transition relation $\widehat{R}_k$ is respected by this encoding.

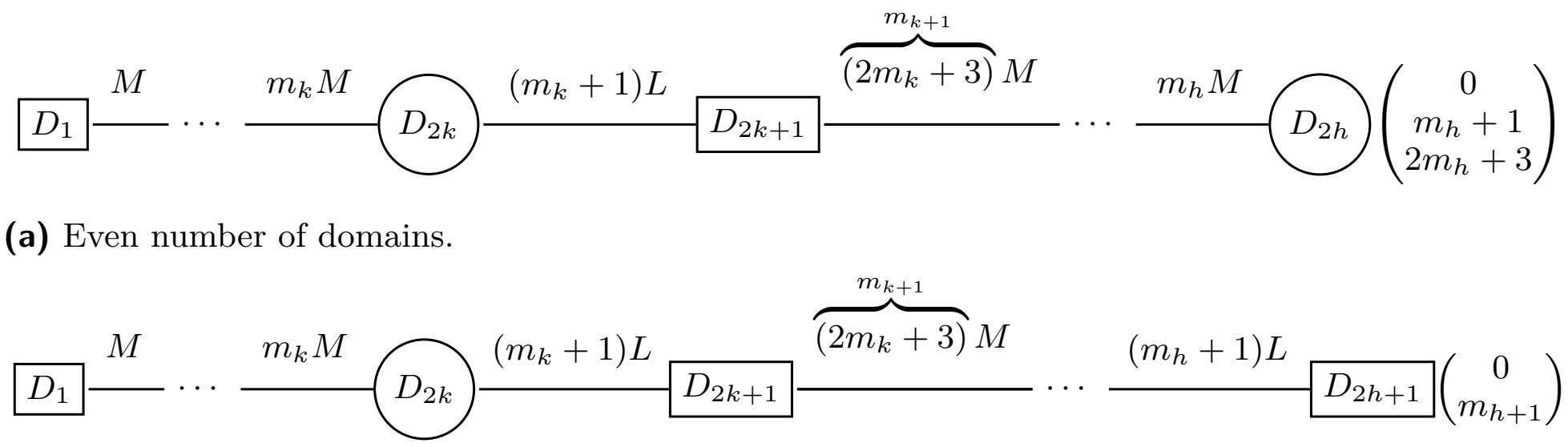

**(a)** Even number of domains.

**(b)** Odd number of domains.

**Figure 1** Constraint graphs for 2-by-3 VCSP when there are an (a) even number of domains and when there are an (b) odd number of domains. Constraints $L$ and $M$ are given in Equation (7). Boxes represend domains with two values, and circles represent domains with three values.

## 4 Pair of 2-by-3 Constraints with Long Ordered Ascent

To build an ordered ascent of exponential length, we consider a VCSP on $n$ domains $D_1 \times D_2 \times \cdots \times D_n$, where the odd domains have size 2 (i.e. $D_1, D_{2k+1} = \{A, B\}$), and the even domains have size 3 (i.e. $D_{2k} = \{A, B, C\}$. For the odd domains, the transition relation is simply given by $\{(A, B)\}$. For the even domains, the transition relation is given by $\{(A, B), (B, C)\}$. In other words, the values are only allowed to transition between $A$-and-$B$, and between $B$-and-$C$. Transitions between $A$-and-$C$ are not allowed.

We arrange the domains into a path, where each pair of consecutive domains have a binary constraint between them. The $n$-th domain gets a unary constraint inspired by the relevant binary constraint. The binary constraints are different weights of the following:

$$L = \begin{pmatrix} 0 & 2 \\ 1 & 1 \\ 2 & 0 \end{pmatrix} \begin{matrix} A \\ B \\ C \end{matrix} \ (\text{columns } A, B), \quad M = \begin{pmatrix} 0 & 1 & 0 \\ 1 & 0 & 1 \end{pmatrix} \begin{matrix} A \\ B \end{matrix} \ (\text{columns } A, B, C) \tag{7}$$

We recursively define weights for these constraints by setting $m_1 = 1$, and $m_{k+1} = 2m_k + 3$. We can solve this recurrence relation to get $m_k = 2^{k+1} - 3$. To $x_1$ and $x_2$ we assign constraint $M$. Between $D_{2k}$ and $D_{2k+1}$ we set constraint $(m_k + 1)L$. Between $D_{2k+1}$ and $D_{2(k+1)}$ we set constraint $m_{k+1}M$. Finally, if $n = 2h$, we assign unary constraint $(m_h + 1)L(-, 0)$ to $D_n$. If $n = 2h + 1$, we assign unary constraint $m_{h+1}M(-, 0)$ to $D_n$. These cases are shown in Figure 1.

Note that $m_{k+1} = 2(m_k + 1) + 1$. Combined with the transition relation, this implies that any fitness gain from the right constraint will always outweigh the fitness loss from the left. The maximal fitness value that can be attained by an assignment of length $n$ is given by

$$f_{\max}(n) := \begin{cases} \sum_{i=1}^{h} (3m_i + 2) = 3 \cdot 2^{h+2} - 7h - 12, & n = 2h \text{ for some } h > 0 \\ f_{\max}(2h) + m_{h+1} = 2^{h+4} - 7h - 15, & n = 2h + 1 \text{ for some } h \geq 0. \end{cases} \tag{8}$$

This fitness value is attained by the assignment $BABA \ldots ABABC$ for $n = 2h$ and the assignment $BABA \ldots ABAB$ for $n = 2h + 1$.

We now prove that there is an ascent that takes on all fitness values from 0 to $f_{\max}$ and is thus exponentially long in the number of variables.

▶ **Proposition 11.** *Let $n \in \mathbb{N}$, and let $D_1 \times D_2 \times \cdots \times D_{n-1} \times D_n$ be the search space given by setting $D_{2k} = \{A, B, C\}$ and $D_{2k+1} = \{A, B\}$, with only transitions between $A$-and-$B$ and $B$-and-$C$ being allowed. Let $p = (x^t)_{t=0}^T$ be the $<$-ordered ascent on the VCSP with constraints from Figure 1, given by starting at $x^0 = A^n$. Then, $p$ has length $f_{\max}(n) \geq 3 \cdot 2^{\lfloor \frac{n}{2} \rfloor} - O(n)$.*

**Proof.** We will show that for any assignment on $n$ variables whose fitness is below $f_{\max}(n)$, there is a fitness increasing move. Moreover, the fitness increasing move on the least index where such a move is possible yields a fitness increase of 1. Since $A^n$ has fitness 0, it follows from these facts that $p$ has length $f_{\max}(n)$.

Let $x \in D_1 \times \cdots \times D_n$ be an assignment on $n$ variables and suppose $f(x) < f_{\max}(n)$. Then, there must be some position $k \leq n$ such that the constraint between $x_k$ and $x_{k+1}$ is not saturated (or in the case that $k = n$, then the unary constraint on $x_n$ is not saturated). Consider the least $k$ for which is the case.

Firstly, consider the case where $k = 1$. Then, we must have $M(x_1, x_2) = 0$. By considering $M$, we see that we can change $x_1$ to obtain a increase of 1.

Next, consider the case where $k = 2l$. We must have that $M(x_{k-1}, x_k) = 1$, since by assumption, this constraint is saturated. Recall that the transition relations of our domains are such that we may only transition between $A$-and-$B$, and between $B$-and-$C$. By considering the constraints, we see that if $x_k = B$, one of the two options yields a increase of $m_l + 1$ from the right constraint (which is an $L$-constraint with weight $m_l + 1$), while the other yields a decrease of $m_l + 1$ from right constraint. If $x_k = A$ or $x_k = C$, then the fact that $M(x_{k-1}, x_k) = 1$, in conjunction with $L(x_k, x_{k+1})$ not being saturated, ensures that changing $x_k$ to $B$ is guaranteed to yield an increase of $m_l + 1$ from the right constraint. Moreover, we are in any case guaranteed to lose $m_l$ from the left constraint by changing $x_k$. Thus, the fitness-improving change of $x_k$ yields a net fitness increase of 1 by changing $x_k$.

Finally, consider the case where $k = 2l + 1$. We must have that $L(x_{k-1}, x_k) = 2$, and $M(x_k, x_{k+1}) = 0$. By considering the constraints, we see that the only possible change for $x_k$ yields a fitness increase of $2m_l + 3 = 2(m_l + 1) + 1$, from the right constraint, while it yields a fitness loss of $2(m_l + 1)$ from the left constraint. Thus, this step again yields a net fitness increase of 1. In the cases where $k = n$, the above arguments hold when we fix $x_{k+1} = A$. ◀

## 5 Steepest Ascent Simulation of 2-by-3 Ordered Ascent

Since the exponentially long ascent described in Section 4 is ordered, we can construct a steepest ascent that simulates it by applying the technique introduced in Section 3. This yields a steepest ascent on an expanded VCSP with alternating 5-state and 3-state domains, and constraints with arity at most 3. The construction has three steps: (i) we define constraints that implement Equation (2), then we define constraints that implement Equation (3) for (ii) odd domains and for (iii) even domains.

Our expanded domains are given by $\widehat{D}_1, \widehat{D}_{2k+1} = \{A, B, \sigma_{AB}\}$ for the odd domains, and $\widehat{D}_{2k} = \{A, B, C, \sigma_{AB}, \sigma_{BC}\}$ for the even domains. In order to obtain the required fitness value for our main states, we add expanded versions $\widehat{L}$ and $\widehat{M}$ from Equation (7). These expanded constraints are still binary, and are given by firstly setting $\widehat{L}_{uv} = L_{uv}$ and $\widehat{M}_{vu} = M_{vu}$ for all $u \in \{A, B, C\}$ and $v \in \{A, B\}$. All other entries in $\widehat{L}$ and $\widehat{M}$, for which at least one of the indices is an intermediate state, are set to 0.

We place these constraints into a path as in Section 4, with new weight $2n + 1$ times the original weight. For any assignment $x$ containing only main states, these expanded constraints yield $\widehat{f}(x) = (2n + 1)f(x)$, which is Equation (2), as desired.

We need to define constraints that ensure that Equation (3) holds for neighbourhoods of odd domains. This will be done by introducing a ternary constraint $\widehat{T}$ that we call the minimisation constraint, as well as unary constraint $\widehat{U}$.

Suppose intermediate state $\sigma_{AB}$ is at odd position $k = 2l+1$. We assume $u = x_{k-1}$ and $v = x_{k+1}$ are main states. Equation (3) requires our restricted fitness function $\widehat{f}_k$ to satisfy:

$$\widehat{f}_k(\sigma_{AB}; u, v) = n - k + 1 + (2n+1) \min_{h \in \{A,B\}} \{f_k(h; u, v)\} \tag{9}$$

$$= n - k + 1 + (2n+1) \min \begin{cases} (m_l + 1) \cdot L(u, A) + m_{l+1} \cdot M(A, v) \\ (m_l + 1) \cdot L(u, B) + m_{l+1} \cdot M(B, v) \end{cases} \tag{10}$$

$$= n - k + 1 + (2n+1)(m_l + 1) \cdot P(u, v) \tag{11}$$

where $P = \begin{pmatrix} 0 & 2 & 0 \\ 1 & 1 & 1 \\ 2 & 0 & 2 \end{pmatrix}$ (columns $A, B, C$; rows $A, B, C$). $P$ specifies the non-zero part of the minimisation constraint $\widehat{T}$, by setting $\widehat{T}_{uwv} = P_{uv}$ for $u, v \in \{A, B, C\}$ and $w = \sigma_{AB}$. All other entries in $\widehat{T}$ (i.e. those for $w$ is a main state, or at least one of $u$ and $v$ is an intermediate state) are set to 0. In order to get the $+n - k + 1$ term from Equation (3), we need a unary constraint $\widehat{U}^\top = \begin{pmatrix} 0 & 0 & 1 \end{pmatrix}$ (columns $A, B, \sigma_{AB}$). For odd $k = 2l+1$, we assign $\widehat{U}$ with weight $n - k + 1$ to $\widehat{D}_k$. We assign ternary constraint $\widehat{T}$ with weight $(2n+1)(m_l+1)$ to $\widehat{D}_k$, $\widehat{D}_{k-1}$ and $\widehat{D}_{k+1}$.

Now, suppose that $x$ is an assignment with exactly one intermediate symbol $\sigma_{AB}$ at odd position $k = 2l+1$. For any two adjacent main states in $x$, the binary constraint between them is given by the binary constraint from the original VCSP, multiplied by a factor $2n+1$. Moreover, $\widehat{T}$ is 0 when there are two adjacent main states among its three indices. Together $\widehat{T}$ and $\widehat{U}$ ensure that for our single intermediate state $\sigma_{AB}$ at position $k$, Equation (10) holds. This yields $\widehat{f}(x) = n - k + 1 + (2n+1) \min_{h \in A,B} \{f(x[k:h])\}$ – the desired Equation (3).

Next, consider intermediate symbol $w \in \{\sigma_{AB}, \sigma_{BC}\}$ at even position $k = 2l$. Assume $u = x_{k-1}$ and $v = x_{k+1}$ are main states. Equation (3) requires:

$$\widehat{f}_k(\sigma_{AB}; u, v) = n - k + 1 + (2n+1) \min_{h \in \{A,B\}} \{f_k(x[k:h])\} \tag{12}$$

$$= n - k + 1 + (2n+1) \min \begin{cases} m_l \cdot M(u, A) + (m_l + 1) \cdot L(A, v) \\ m_l \cdot M(u, B) + (m_l + 1) \cdot L(B, v) \end{cases} \tag{13}$$

$$= n - k + 1 + (2n+1) \cdot Q^l(u, v) \tag{14}$$

$$\widehat{f}_k(\sigma_{BC}; u, v) = n - k + 1 + (2n+1) \min_{h \in \{B,C\}} \{f_k(x[k:h])\} \tag{15}$$

$$= n - k + 1 + (2n+1) \min \begin{cases} m_l \cdot M(u, B) + (m_l + 1) \cdot L(B, v) \\ m_l \cdot M(u, C) + (m_l + 1) \cdot L(C, v) \end{cases} \tag{16}$$

$$= n - k + 1 + (2n+1) \cdot R^l(u, v) \tag{17}$$

where $Q^l = \begin{pmatrix} 0 & 2m_l + 1 \\ m_l & m_l + 1 \end{pmatrix}$ (columns $A, B$; rows $A, B$) and $R^l = \begin{pmatrix} 2m_l + 1 & 0 \\ m_l + 1 & m_l \end{pmatrix}$ (columns $A, B$; rows $A, B$). The matrices $Q^l$ and $R^l$ specify the non-zero part of the ternary minimisation constraint $\widehat{S}^l$, by setting $\widehat{S}^l_{uwv} = Q^l_{uv}$ for $w = \sigma_{AB}$, and $\widehat{S}^l_{uwv} = R^l_{uv}$ for $w = \sigma_{BC}$. All other entries in $\widehat{S}^l$ (i.e. those for $w$ is a main state, or at least one of $u$ and $v$ is an intermediate state) are set to 0. Furthermore, in order to get the first summands of Equation (3), we need a unary constraint $\widehat{V}^\top = \begin{pmatrix} \overset{A}{0} & \overset{B}{0} & \overset{C}{0} & \overset{\sigma_{AB}}{1} & \overset{\sigma_{BC}}{1} \end{pmatrix}$. For any even $k = 2l$, we assign constraint $\widehat{V}$ with weight $n - k + 1$ to $\widehat{D}_k$. We also assign the ternary constraint $S^l$ with weight $2n+1$ to $\widehat{D}_k$ and its neighbouring domains $\widehat{D}_{k-1}$ and $\widehat{D}_{k+1}$.[3]

Now, suppose that $x$ is an assignment with exactly one intermediate symbol $w \in \{\sigma_{AB}, \sigma_{BC}\}$ at even position $k = 2l$. Note that constraints $\widehat{T}$ and $\widehat{U}$ do not add any fitness value for this assignment. Thus, in the same manner as before, the binary constraints $\widehat{M}$ and $\widehat{L}$, together with $\widehat{S}$ and $\widehat{V}$ yield the desired $\widehat{f}(x) = k + (2n+1)f \min_{h \in A,B}(\{f(x[k:h])\}$.

In order to apply Theorem 8, it only remains to show that the inequality from Equation (4) holds. By inspecting $\widehat{S}^l$ and $\widehat{T}$ however, it is clear that this is the case. Specifically, we have equality in the case of two non-adjacent intermediate states, and strict inequality in the case of adjacent intermediate states.

## 6 Low-arity Boolean Encoding for 3-by-5 Steepest Ascent

In order to turn the alternating 3-state and 5-state VCSP from Section 5 into a Boolean VCSP, we encode the expanded domains using Boolean variables as in Section 3.3. The 3-state domain is encoded by two boolean variables with $A = 10$, $B = 01$ and $\sigma_{AB} = 11$. The 5-state domain is encoded using three variables with $A = 100$, $B = 010$, $C = 001$, $\sigma_{AB} = 110$ and $\sigma_{BC} = 011$. We write $\widehat{\widehat{D}}_k$ for an encoded domain and $\widehat{\widehat{f}}$ for our new fitness function.[4]

Note that the constraints can be encoded by taking the value of the original constraint for any encoded states and 0 for strings that don't encode any states. In the encoded VCSP, the arity of the constraints defined in Section 5 increases. The $\widehat{S}^l$-constraint, which was a ternary constraint on a 3-state, a 5-state and another 3-state domain, turns into a constraint with arity $2 + 3 + 2 = 7$ in the encoded VCSP. The $\widehat{T}$-constraint, which was a ternary constraint on a 5-state, a 3-state, and another 5-state domain, turns into a constraint with arity $3 + 2 + 3 = 8$. We will use some tricks to reduce this arity to 5 in both cases. The resulting full encoded VCSP is shown in Figure 2.

First, we look at a lower arity implementation of the minimisation constraint $\widehat{T}$ for 3-state domain neighbourhoods, from Section 5. Suppose that we have an intermediate state $\sigma_{AB}$ at odd position $k = 2l + 1$. We assume that $x_{k-1}$ and $x_{k+1}$ are main states, represented by $u$ and $v$ respectively. Recall that the non-zero part of $\widehat{T}$ is given by $\widehat{T}_{*\sigma_{AB}*} = \begin{pmatrix} \overset{A}{0} & \overset{B}{2} & \overset{C}{0} \\ 1 & 1 & 1 \\ 2 & 0 & 2 \end{pmatrix} \begin{matrix} A \\ B \\ C \end{matrix}$.[5]

Instead of encoding $\sigma_{AB}$ with the single string 11, we can let both 00 and 11 perform part of

---

Note that introducing these new constraints does not impact Equation (3), since $\widehat{V}$ adds fitness value 0 for main states, and $S_{uwv}$ adds fitness value 0 if $w$ is a main state or $u$ or $v$ is an intermediate symbol (which are the only situations occurring in Equation (3)).

$\widehat{\widehat{f}}$ takes the same values as our old fitness function $\widehat{f}$, but has a new domain.

One could try to decompose $\widehat{T}$ into two arity-5 constraints between $\widehat{\widehat{D}}_{k-1}$-and-$\widehat{\widehat{D}}_k$ and one between $\widehat{\widehat{D}}_k$-and-$\widehat{\widehat{D}}_{k+1}$. This is not possible if $\sigma_{AB}$ is encoded by a single string (see Appendix A).

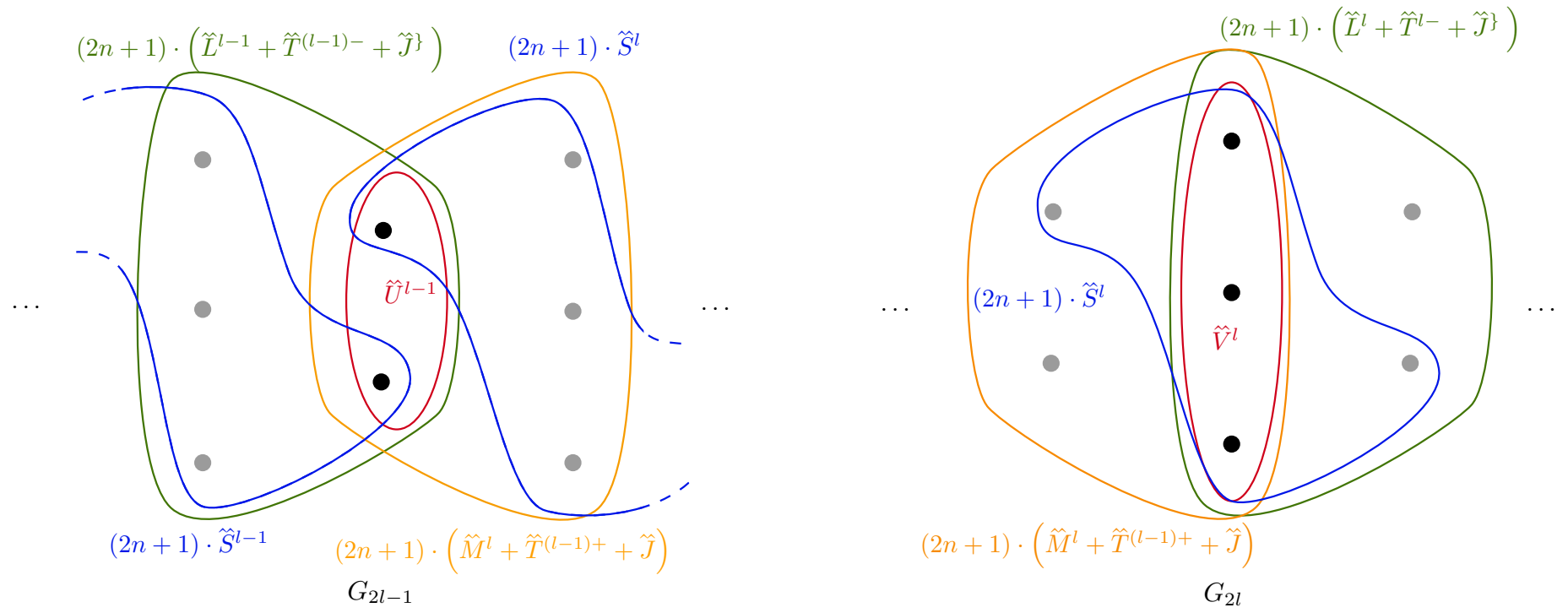

**(a)** **(b)**

**Figure 2** The final VCSP contains $5n$ domains, divided alternatingly into collections $G$ , $G_{\ l}$ of 2 domains and collections $G_{\ l}$ of 3 domains. There are eight different types of constraints $\hat{\hat{M}}^l$, $\hat{\hat{L}}^l$, $\hat{\hat{T}}^{l-}$, $\hat{\hat{T}}^{l}$ , $\hat{\hat{S}}^l$, $\hat{\hat{J}}$, $\hat{\hat{U}}^l$, and $\hat{\hat{V}}^l$. Their values can be found in Equations (23)-(30). These constraints are arranged in a path of repeating chunks. These chunks are shown for (a) odd collections and (b) even collections. As exceptions, we do not include a $\hat{\hat{T}}$ constraint between the first two collections $G$ and $G$ . Furthermore, for the final collection $G_n$, we pretend the pattern continues, but that the values for collections beyond $G_n$ are fixed to 100 for even collections and 10 for odd collections. The steepest ascent starting from assignment $10100^{\lfloor \frac{n}{2} \rfloor}$ (with an extra 10 at the end in case $n$ is odd), has length at least $3 \cdot 2^{\lfloor \frac{n}{2} \rfloor \ \ 1} - O(n)$ (see Proposition 11 and Theorem 8 ). Furthemore, the constraint graph has pathwidth 4. To see this, take for the path decomposition the scopes of the $\hat{\hat{M}}^l$ constraints, the scopes of the $\hat{\hat{S}}^l$ constraints, and the scopes of the $\hat{\hat{L}}^l$ constraints as the bins, and put these bins in the path $\cdots$-$\hat{\hat{M}}^l$-$\hat{\hat{S}}^l$-$\hat{\hat{L}}^l$-$\hat{\hat{M}}^l$ -$\cdots$.

the role of encoding $\sigma_{AB}$. Importantly, this encoding still agrees with out transition relation, since 00 and 11 two bit-flips away from one another, and a single bit flip away from 10 and 01. We now have to set up the fitness values in such a way that we can transition through either 00 or 11. In order for this to work, we require that $\max_{s\in\{00,11\}} \hat{\hat{f}}_k(s;u,v) = \hat{f}_k(\sigma_{AB};u,v)$ for all $u, v$. We may view $u$ and $v$ as vectors, writing them as vectors with $\vec{A} = \overset{\begin{matrix}A & B & C\end{matrix}}{\begin{pmatrix}1 & 0 & 0\end{pmatrix}}$, $\vec{B} = \overset{\begin{matrix}A & B & C\end{matrix}}{\begin{pmatrix}0 & 1 & 0\end{pmatrix}}$ and $\vec{C} = \overset{\begin{matrix}A & B & C\end{matrix}}{\begin{pmatrix}0 & 0 & 1\end{pmatrix}}$. We can achieve our desired property with:

$$\hat{\hat{f}}_k(00;u,v) = n-k+1+(2n+1)(m_l+1)\vec{u}\begin{pmatrix}0 & -2 & 0\\ 1 & -1 & 1\\ 2 & 0 & 2\end{pmatrix}\vec{v} \tag{18}$$

$$= n-k+1+(2n+1)(m_l+1)\vec{u}\left(\begin{pmatrix}0\\1\\2\end{pmatrix}\begin{pmatrix}1 & 1 & 1\end{pmatrix} + \begin{pmatrix}1\\1\\1\end{pmatrix}\begin{pmatrix}0 & -2 & 0\end{pmatrix}\right)\vec{v}, \tag{19}$$

$$\hat{\hat{f}}_k(11;u,v) = n-k+1+(2n+1)(m_l+1)\vec{u}\begin{pmatrix}0 & 2 & 0\\ -1 & 1 & -1\\ -2 & 0 & -2\end{pmatrix}\vec{v} \tag{20}$$

$$= n-k+1+(2n+1)(m_l+1)\vec{u}\left(\begin{pmatrix}2\\1\\0\end{pmatrix}\begin{pmatrix}1 & 1 & 1\end{pmatrix} + \begin{pmatrix}1\\1\\1\end{pmatrix}\begin{pmatrix}-2 & 0 & -2\end{pmatrix}\right)\vec{v}. \tag{21}$$

Note that $\widehat{T}_{*\sigma_{AB}*}$ is the element-wise maximum of the two $3 \times 3$ matrices above. From this, we obtain our desired arity-5 constraints $\widehat{\widehat{T}}^{l-}$ and $\widehat{\widehat{T}}^{l+}$ between $\widehat{\widehat{D}}_{k-1}$ and $\widehat{\widehat{D}}_k$, and between $\widehat{\widehat{D}}_k$ and $\widehat{\widehat{D}}_{k+1}$. The non-zero part of $\widehat{\widehat{T}}^{l-}$ and $\widehat{\widehat{T}}^{l+}$ are given by

$$\widehat{\widehat{T}}^{l-} = \begin{array}{c} \begin{matrix} \scriptstyle 00 & \scriptstyle 11 \end{matrix} \\ \begin{pmatrix} 0 & 2(m_l+1) \\ m_l+1 & m_l+1 \\ 2(m_l+1) & 0 \end{pmatrix} \end{array} \begin{matrix} \scriptstyle 100 \\ \scriptstyle 010 \\ \scriptstyle 001 \end{matrix} \quad \text{and } \widehat{\widehat{T}}^{l+} = -2(m_l+1) \cdot \begin{array}{c} \begin{matrix} \scriptstyle 100 & \scriptstyle 010 & \scriptstyle 001 \end{matrix} \\ \begin{pmatrix} 0 & 1 & 0 \\ 1 & 0 & 1 \end{pmatrix} \end{array} \begin{matrix} \scriptstyle 00 \\ \scriptstyle 11 \end{matrix} . \tag{22}$$

Next, we look at a lower arity implementation of the minimisation constraint $\widehat{S}^l$. Suppose that we have an intermediate state $\sigma_{AB}$ at even position $k = 2l$. We assume that $x_{k\ 1}$ and $x_{k+1}$ are main states, represented by $u$ and $v$ respectively. Recall that the non-zero parts of $\widehat{S}^l$ are given by $\widehat{S}^l{}_{\sigma_{AB}} = \begin{array}{c} \begin{matrix} \scriptstyle A & \scriptstyle B \end{matrix} \\ \begin{pmatrix} 0 & 2m_l+1 \\ m_l & m_l+1 \end{pmatrix} \end{array} \begin{matrix} \scriptstyle A \\ \scriptstyle B \end{matrix}$ and $\widehat{S}^l{}_{\sigma_{BC}} = \begin{array}{c} \begin{matrix} \scriptstyle A & \scriptstyle B \end{matrix} \\ \begin{pmatrix} 2m_l+1 & 0 \\ m_l+1 & m_l \end{pmatrix} \end{array} \begin{matrix} \scriptstyle A \\ \scriptstyle B \end{matrix}$ .

This time we will use a different trick to lower the arity of the encoded constraint from 7 to 5. Note that only one domain can enter transition. Since we are considering domain $D_k$ entering transition, we can assume that $u$ and $w$ are main states. This implies that we do not need to look at both bits of $u$'s and $w$'s representation to know their value. We can just look at the right bit of $u$ and the left bit of $v$. This reduces the arity of the constraint to 5. We get a new constraint $\widehat{\widehat{S}}^l$ whose scope consists of right bit of the encoding of $x_{k\ 1}$, three bits that encode $x_k$, and left bit of the encoding of $x_{k+1}$. Non-zero parts of $\widehat{\widehat{S}}^l$ are

$\widehat{\widehat{S}}^l{}_{110} = \begin{array}{c} \begin{matrix} \scriptstyle 1 & \scriptstyle 0 \end{matrix} \\ \begin{pmatrix} 0 & 2m_l+1 \\ m_l & m_l+1 \end{pmatrix} \end{array} \begin{matrix} \scriptstyle 0 \\ \scriptstyle 1 \end{matrix}$ and $\widehat{\widehat{S}}^l{}_{011} = \begin{array}{c} \begin{matrix} \scriptstyle 1 & \scriptstyle 0 \end{matrix} \\ \begin{pmatrix} 2m_l+1 & 0 \\ m_l+1 & m_l \end{pmatrix} \end{array} \begin{matrix} \scriptstyle 0 \\ \scriptstyle 1 \end{matrix}$ .

By restricting our perspective to only a single bit of the representation of $x_{k\ 1}$ and $x_{k+1}$, we can no longer distinguish whether $x_{k\ 1}$ and $x_{k+1}$ are main states, or whether they are intermediate states. Through this, we may inadvertently violate Equation (4). To remedy this, we introduce an arity-5 constraint $\widehat{J}$ that penalises the occurrence of two adjacent intermediate states by setting the non-zero part of $\widehat{J}$ as $-f_{\max}(n) \cdot \begin{array}{c} \begin{matrix} \scriptstyle 110 & \scriptstyle 011 \end{matrix} \\ \begin{pmatrix} 1 & 1 \\ 1 & 1 \end{pmatrix} \end{array} \begin{matrix} \scriptstyle 00 \\ \scriptstyle 11 \end{matrix}$ where $n$ is the number of domains in the original VCSP and $f_{\max}(n)$ is the maximal fitness from Equation (8) of the original 2-by-3 VCSP. The magnitude of this negative value is always larger than the magnitude of any fitness value assigned by $S^l$ for any $l$.

We have now constructed a VCSP with Boolean domains. We arrange these domains alternatingly into collections $G_{2l}$ consisting of 3 domains and $G_{2l+1}$ of 2 domains. The VCSP has eight different types of constraints which we list explicitly. We have an arity-5 constraint $\widehat{\widehat{M}}^l$ on collections $G_{2l\ 1}$ and $G_{2l}$:

$$\widehat{\widehat{M}}^l = \begin{array}{c} \begin{matrix} \scriptstyle 100 & \scriptstyle 010 & \scriptstyle 001 & \scriptstyle 110 & \scriptstyle 101 & \scriptstyle 011 & \scriptstyle 000 & \scriptstyle 111 \end{matrix} \\ \begin{pmatrix} 0 & m_l & 0 & 0 & 0 & 0 & 0 & 0 \\ m_l & 0 & m_l & 0 & 0 & 0 & 0 & 0 \\ 0 & 0 & 0 & 0 & 0 & 0 & 0 & 0 \\ 0 & 0 & 0 & 0 & 0 & 0 & 0 & 0 \end{pmatrix} \end{array} \begin{matrix} \scriptstyle 10 \\ \scriptstyle 01 \\ \scriptstyle 00 \\ \scriptstyle 11 \end{matrix} \tag{23}$$

and an an arity-5 constraint $\widetilde{L}^l$ on collections $G_{2l}$ and $G_{2l+1}$, given by:

$$\widehat{\widehat{L}}^l = \begin{array}{cc} \begin{array}{cccc} \scriptstyle 10 & \scriptstyle 01 & \scriptstyle 00 & \scriptstyle 11 \end{array} & \\ \begin{pmatrix} 0 & 2(m_l+1) & 0 & 0 \\ m_l+1 & m_l+1 & 0 & 0 \\ 2(m_l+1) & 0 & 0 & 0 \\ 0 & 0 & 0 & 0 \\ 0 & 0 & 0 & 0 \\ 0 & 0 & 0 & 0 \\ 0 & 0 & 0 & 0 \\ 0 & 0 & 0 & 0 \end{pmatrix} & \begin{array}{c} \scriptstyle 100 \\ \scriptstyle 010 \\ \scriptstyle 001 \\ \scriptstyle 110 \\ \scriptstyle 101 \\ \scriptstyle 011 \\ \scriptstyle 000 \\ \scriptstyle 111 \end{array} \end{array} \tag{24}$$

Then, we have our (decomposed) minimisation constraints $T^{l-}$ and $T^{l+}$, on collections $G_{2l}$ and $G_{2l+1}$, and collections $G_{2l+1}$ and $G_{2(l+1)}$, respectively. These constraints are given by:

$$\widehat{\widehat{T}}^{l-} = \begin{array}{cc} \begin{array}{cccc} \scriptstyle 10 & \scriptstyle 01 & \scriptstyle 00 & \scriptstyle 11 \end{array} & \\ \begin{pmatrix} 0 & 0 & 0 & 2(m_l+1) \\ 0 & 0 & m_l+1 & m_l+1 \\ 0 & 0 & 2(m_l+1) & 0 \\ 0 & 0 & 0 & 0 \\ 0 & 0 & 0 & 0 \\ 0 & 0 & 0 & 0 \\ 0 & 0 & 0 & 0 \\ 0 & 0 & 0 & 0 \end{pmatrix} & \begin{array}{c} \scriptstyle 100 \\ \scriptstyle 010 \\ \scriptstyle 001 \\ \scriptstyle 110 \\ \scriptstyle 101 \\ \scriptstyle 011 \\ \scriptstyle 000 \\ \scriptstyle 111 \end{array} \end{array} \tag{25}$$

$$\widehat{\widehat{T}}^{l+} = -2(m_l+1) \cdot \begin{array}{cc} \begin{array}{cccccccc} \scriptstyle 100 & \scriptstyle 010 & \scriptstyle 001 & \scriptstyle 110 & \scriptstyle 101 & \scriptstyle 011 & \scriptstyle 000 & \scriptstyle 111 \end{array} & \\ \begin{pmatrix} 0 & 0 & 0 & 0 & 0 & 0 & 0 & 0 \\ 0 & 0 & 0 & 0 & 0 & 0 & 0 & 0 \\ 0 & 1 & 0 & 0 & 0 & 0 & 0 & 0 \\ 1 & 0 & 1 & 0 & 0 & 0 & 0 & 0 \end{pmatrix} & \begin{array}{c} \scriptstyle 10 \\ \scriptstyle 01 \\ \scriptstyle 00 \\ \scriptstyle 11 \end{array} \end{array} \tag{26}$$

The next minimisation constraint is the arity-5 constraint $\widehat{\widehat{S}}^l$ with scope consisting of a single domain in $G_{2l-1}$, all three domains in $G_{2l}$ and one more domain in $G_{2l+1}$. The single domains are selected such that the scopes of $\widehat{\widehat{S}}^l$ and $\widehat{\widehat{S}}^{l+1}$ do not overlap. $\widehat{\widehat{S}}^l$ is given by:

$$\widehat{\widehat{S}}^l = \begin{array}{cc} \begin{array}{cccc} \scriptstyle (0,0) & \scriptstyle (1,0) & \scriptstyle (0,1) & \scriptstyle (1,1) \end{array} & \\ \begin{pmatrix} 0 & 0 & 0 & 0 \\ 0 & 0 & 0 & 0 \\ 0 & 0 & 0 & 0 \\ 2m_l+1 & m_l+1 & 0 & m_l \\ 0 & 0 & 0 & 0 \\ 0 & m_l & 2m_l+1 & m_l+1 \\ 0 & 0 & 0 & 0 \\ 0 & 0 & 0 & 0 \end{pmatrix} & \begin{array}{c} \scriptstyle 100 \\ \scriptstyle 010 \\ \scriptstyle 001 \\ \scriptstyle 110 \\ \scriptstyle 101 \\ \scriptstyle 011 \\ \scriptstyle 000 \\ \scriptstyle 111 \end{array} \end{array} \tag{27}$$

where the $(u, v)$ for the column indices takes $u$ and $v$ as the values for respective single domains in $G_{2l-1}$ and $G_{2l+1}$. In order to ensure that this constraint does not inadvertently allow adjacent intermediate states, we introduce an arity-5 constraint $\hat{\hat{J}}$ on adjacent collections:

$$\hat{\hat{J}} = -f_{\max}(n) \cdot \begin{pmatrix} \overset{100}{0} & \overset{010}{0} & \overset{001}{0} & \overset{110}{0} & \overset{101}{0} & \overset{011}{0} & \overset{000}{0} & \overset{111}{0} \\ 0 & 0 & 0 & 0 & 0 & 0 & 0 & 0 \\ 0 & 0 & 0 & 1 & 0 & 1 & 0 & 0 \\ 0 & 0 & 0 & 1 & 0 & 1 & 0 & 0 \end{pmatrix} \begin{matrix} {\scriptstyle 10} \\ {\scriptstyle 01} \\ {\scriptstyle 00} \\ {\scriptstyle 11} \end{matrix} \tag{28}$$

where $n$ is the number of collections, and $f_{\max}(n)$ is as defined in Equation (8).

Finally, we have arity-3 constraint $V^l$ on $G_{2l}$ and arity-2 constraint $U^l$ on $G_{2l+1}$:

$$(\hat{\hat{U}}^l)^\top = \begin{pmatrix} \overset{10}{0} & \overset{01}{0} & \overset{00}{n - 2l} & \overset{11}{n - 2l} \end{pmatrix} \tag{29}$$

$$(\hat{\hat{V}}^l)^\top = \begin{pmatrix} \overset{100}{0} & \overset{010}{0} & \overset{001}{0} & \overset{110}{n - 2l + 1} & \overset{101}{0} & \overset{011}{n - 2l + 1} & \overset{000}{0} & \overset{111}{0} \end{pmatrix} \tag{30}$$

We arrange these constraints into a path of repeating chunks between the collections, as shown in Figure 2.

## 7 Summary, Future Directions, and Biological Evolution

In this paper, we presented a sequence of three constructions with each improving on the state-of-the-art for the "simplest" VCSP with some desired property of intractability. In Section 4, we presented a binary VCSP with a path as its constraint graph and domains alternating in size between two-state and three-state. In Proposition 11, we showed that this 2-by-3 VCSP has an exponential ascent. This a simplification over Kaznatcheev, Cohen and Jeavons [12]'s simplest example of an exponential ascent from a path-structured VCSP with all domains of size 3. Our example also has the added benefit over prior work of the exponential ascent being an ordered-ascent. The ascent being ordered allows us to apply our general padding technique from Section 3 to create a ternary VCSP with domains alternating between size 3 and 5 in Section 5. It then follows from Theorem 8 that this 3-by-5 ternary VCSP produces a fitness landscapes with exponential steepest ascents. This could be viewed as a simplification over the binary VCSP with domains of size 10 implicit in Cohen et al. [3]'s construction of exponential steepest ascents. Finally, in Section 6, we encoded the 3-by-5 VCSP with Boolean domains to construct a Boolean VCSP with a constraint graph of pathwidth 4 that produce a fitness landscape with an exponentially long steepest ascent. This is an improvement over the pathwidth 7 of the best known prior construction [3].

Our final construction means that Boolean VCSPs of pathwidth 4 are intractable for local search by steepest ascent. Since our graph also has treewidth 4 this means that Boolean VCSPs of treewidth 4 are also intractable for steepest ascent. For tractability, Kaznatcheev, Cohen and Jeavons [12] have shown that all ascents – and thus in particular the steepest ascent – have at most quadratic length when the constraint graph is a tree, i.e. has treewidth 1. This leaves a gap between treewidth 1 and treewidth 4 for which the status of steepest ascent for finding local maxima in Boolean VCSPs remains unknown. Our current best guess at the exact location of the tractability boundary for steepest ascent is at pathwidth 2:

▶ **Conjecture 12.** *There exists a polynomial $p(n)$ such that for any Boolean VCSP instance $\mathcal{C}$ on $n$ variables if the constraint graph of $\mathcal{C}$ has pathwidth $\leq 2$, then any steepest ascent in the associated fitness landscape has length at most $p(n)$.*

Of course, the existence of exponential steepest ascents does not mean that all ascents are long. In our construction, it is relatively easy to find a short ascent that violates the steepest ascent condition. In fact, Kaznatcheev [11] has shown that polynomially short ascents to some local solution exist from all initial assignment in fitness landscape from VCSPs of bounded treewidth. More generally, there exist efficient (non-local search) algorithms for finding the *global* maximum in VCSPs of bounded treewidth [1, 6, 2]. However, such global algorithms cannot always be run – especially in cases where the algorithm is actually some natural process and thus we have no (or only partial) control to 'rewrite' the algorithm.

Biological evolution is an important local search algorithm that is set by nature [19, 15, 9]. The intractability of finding local peaks provides an explanations for important features of evolution like it's open-endededness [10, 9]. In this case, we can read ascents as 'adaptive paths' [4] and steepest ascent as a strong-selection weak mutation dynamic that is often studied in evolutionary biology [5, 15]. The VCSP's variables correspond to genetic loci, the valued constraints correspond to gene-interactions, and the constraint graphs of the VCSPs encoding fitness landscapes correspond to gene-interaction networks [18, 9].[6] In this case, finding the 'simplest' VCSPs that have exponential steepest ascents allows us to reason about the minimal conditions for open-endedness in evolution. Thus, our hope is that further progress on local search for VCSPs increases not only our understanding of combinatorial optimization but also of natural processes like biological evolution.

## Acknowledgements

AK would like to thank Dave Cohen and Peter Jeavons for helpful discussions. AK and MvM would also like to thank Daniel Dadush for helpful feedback and questions.

[6] Similar models where the local search algorithm is set by nature exist for social systems studied in business [14, 16] and economics [17]; and for energy-minimization in physical systems.

## A No naive decomposition into arity-5 constraints for T

Let $k > 0$ be an odd integer. One option for reducing the arity of constraint $\widehat{T}$ from Section 5, would be to decompose it into two arity-5 constraints, one between $D_{k-1}$-and-$D_k$ and one between $D_k$-and-$D_{k+1}$. Note that in this case, the value $x_k$ would pick out the column of the $D_{k-1}$-$D_k$-constraint, and the row of the $D_k$-$D_{k+1}$-constraint. Taken together, these two components would need to result in $P$ from Section 5. Thus, if we want to implement this with two arity-5 constraints between we need to show how to implement $P$ as the sum of two rank-1 matrices $Q$:

$$P = \begin{pmatrix} 0 & 1 & 2 \\ 2 & 1 & 0 \\ 0 & 1 & 2 \end{pmatrix} \stackrel{?}{=} \begin{pmatrix} A_1 \\ B_1 \\ C_1 \end{pmatrix} \begin{pmatrix} 1 & 1 & 1 \end{pmatrix} + \begin{pmatrix} 1 \\ 1 \\ 1 \end{pmatrix} \begin{pmatrix} A_2 & B_2 & C_2 \end{pmatrix} \tag{31}$$

$$= \begin{pmatrix} A_1 + A_2 & A_1 + B_2 & A_1 + C_2 \\ B_1 + A_2 & B_1 + B_2 & B_1 + C_2 \\ C_1 + A_2 & C_1 + B_2 & C_1 + C_2 \end{pmatrix} = Q \tag{32}$$

We can see that this is impossible to satisfy because on the side of $Q$ we have:

$$Q_{1,1} + Q_{2,2} = A_1 + A_2 + B_1 + B_2 \tag{33}$$

$$= A_1 + B_2 + B_1 + A_2 = Q_{1,2} + Q_{2,1} \tag{34}$$

but on the side of $P$ we have:

$$P_{1,1} + P_{2,2} = 0 + 1 \tag{35}$$

$$\neq 1 + 2 = P_{1,2} + P_{2,1}. \tag{36}$$

Thus, if the even domains contain only a single intermediate state, we cannot decompose the minisation constriaint $\widehat{T}$ into two arity-5 constraints between $D_{k-1}$-and-$D_k$ and $D_k$-and-$D_{k+1}$.

## References

[1] U. Bertelè and F. Brioschi. 'On non-serial dynamic programming'. In: *Journal of Combinatorial Theory, Series A* 14.2 (1973), pp. 137–148. ISSN: 0097-3165. DOI: https://doi.org/10.1016/0097-3165(73)90016-2.

[2] C. Carbonnel, M. Romero and S. Živný. 'The Complexity of General-Valued Constraint Satisfaction Problems Seen from the Other Side'. In: *SIAM Journal on Computing* 51.1 (2022), pp. 19–69. DOI: 10.1137/19M1250121.

[3] D. A. Cohen et al. 'Steepest ascent can be exponential in bounded treewidth problems'. In: *Operations Research Letters* 48 (3 2020), pp. 217–224.

[4] K. Crona, D. Greene and M. Barlow. 'The peaks and geometry of fitness landscapes.' In: *Journal of Theoretical Biology* 317 (2013), pp. 1–10.

[5] J. H. Gillespie. 'A simple stochastic gene substitution model'. In: *Theoretical population biology* 23.2 (1983), pp. 202–215.

[6] G. Gottlob, G. Greco and F. Scarcello. 'Tractable optimization problems through hypergraph-based structural restrictions'. In: *International Colloquium on Automata, Languages, and Programming*. Springer. 2009, pp. 16–30.

[7] P. Jeavons, A. Krokhin and S. Živný. 'The Complexity of Valued Constraint Satisfaction'. In: *Bulletin of the European Association for Theoretical Computer Science* 113 (2014), pp. 21–55.

[8] D. S. Johnson, C. H. Papadimitriou and M. Yannakakis. 'How easy is local search?' In: *Journal of Computer and System Sciences* 37.1 (1988), pp. 79–100. ISSN: 0022-0000. DOI: https://doi.org/10.1016/0022-0000(88)90046-3.

[9] A. Kaznatcheev. 'Algorithmic Biology of Evolution and Ecology'. PhD thesis. University of Oxford, 2020.

[10] A. Kaznatcheev. 'Computational complexity as an ultimate constraint on evolution'. In: *Genetics* 212.1 (2019), pp. 245–265.

[11] A. Kaznatcheev. 'Local search for valued constraint satisfaction parameterized by treedepth'. In: *ArXiv* (2024).

[12] A. Kaznatcheev, D. A. Cohen and P. Jeavons. 'Representing fitness landscapes by valued constraints to understand the complexity of local search'. In: *Journal of Artificial Intelligence Research* 69 (2020), pp. 1077–1102.

[13] M. W. Krentel. 'On Finding and Verifying Locally Optimal Solutions'. In: *SIAM Journal on Computing* 19.4 (1990), pp. 742–749. DOI: 10.1137/0219052. eprint: https://doi.org/10.1137/0219052.

[14] D. A. Levinthal. 'Adaptation on rugged landscapes'. In: *Management Science* 43.7 (1997), pp. 934–950.

[15] H. A. Orr. 'The genetic theory of adaptation: a brief history.' In: *Nature Reviews. Genetics* 6 (2005), pp. 119–127.

[16] J. W. Rivkin and N. Siggelkow. 'Patterned interactions in complex systems: Implications for exploration'. In: *Management Science* 53.7 (2007), pp. 1068–1085.

[17] T. Roughgarden. 'Computing equilibria: A computational complexity perspective.' In: *Economic Theory* 42 (1 2010), pp. 193–236.

[18] A. Strimbu. 'Simulating Evolution on Fitness Landscapes represented by Valued Constraint Satisfaction Problems'. In: *arXiv:1912.02134* (2019).

[19] S. Wright. 'The roles of mutation, inbreeding, crossbreeding, and selection in evolution'. In: *Proc. of the 6th International Congress on Genetics*. 1932, pp. 355–366.